\documentclass[aps,pre,floats,floatfix,superscriptaddress,twocolumn]{revtex4-1}
\usepackage{latexsym,amsmath}
\usepackage{amsbsy}
\usepackage[pdftex]{graphicx}
\usepackage{amssymb}
\usepackage{epstopdf}
\usepackage[usenames]{color}
\usepackage{float}
\usepackage{hyperref}
\usepackage{graphicx}
\usepackage[normalem]{ulem}

\definecolor{g}{rgb}{0.0, 0.45, 0.45}
\definecolor{orange}{RGB}{255, 155, 0}

\begin{document}

\title{Mercator: uncovering faithful hyperbolic embeddings of complex networks}

\author{Guillermo Garc\'{\i}a-P\'{e}rez}\thanks{Both authors contributed equally to this work}
\affiliation{QTF Centre of Excellence, Turku Centre for Quantum Physics, Department of Physics and Astronomy, University
of Turku, FI-20014 Turun Yliopisto, Finland}
\affiliation{Complex Systems Research Group, Department of Mathematics and Statistics, University
of Turku, FI-20014 Turun Yliopisto, Finland}

\author{Antoine \surname{Allard}}
\thanks{Both authors contributed equally to this work}
\affiliation{D\'epartement  de  physique,  de  g\'enie  physique  et  d’optique, Universit\'e Laval,  Qu\'ebec  (Qu\'ebec),  Canada  G1V 0A6}
\affiliation{Centre de mod\'elisation math\'ematique, Universit\'e  Laval,  Qu\'ebec  (Qu\'ebec),  Canada  G1V 0A6}

\author{M. \'Angeles \surname{Serrano}}
\affiliation{Departament de F\'isica de la Mat\`eria Condensada, Universitat de Barcelona, Mart\'i i Franqu\`es 1, E-08028 Barcelona, Spain}
\affiliation{Universitat de Barcelona Institute of Complex Systems (UBICS), Universitat de Barcelona, Barcelona, Spain}
\affiliation{Instituci\'o Catalana de Recerca i Estudis Avan\c{c}ats (ICREA), Passeig Llu\'is Companys 23, E-08010 Barcelona, Spain}

\author{Mari\'an \surname{Bogu\~n\'a}}
\thanks{Corresponding author: marian.boguna@ub.edu}
\affiliation{Departament de F\'isica de la Mat\`eria Condensada, Universitat de Barcelona, Mart\'i i Franqu\`es 1, E-08028 Barcelona, Spain}
\affiliation{Universitat de Barcelona Institute of Complex Systems (UBICS), Universitat de Barcelona, Barcelona, Spain}

\begin{abstract}
  We introduce Mercator, a reliable embedding method to map real complex networks into their hyperbolic latent geometry. The method assumes that the structure of networks is well described by the Popularity$\times$Similarity $\mathbb{S}^1/\mathbb{H}^2$ static geometric network model, which can accommodate arbitrary degree distributions and reproduces many pivotal properties of real networks, including self-similarity patterns. The algorithm mixes machine learning and maximum likelihood approaches to infer the coordinates of the nodes in the underlying hyperbolic disk with the best matching between the observed network topology and the geometric model. In its fast mode, Mercator uses a model-adjusted machine learning technique performing dimensional reduction to produce a fast and accurate map, whose quality already outperform other embedding algorithms in the literature. In the refined Mercator mode, the fast-mode embedding result is taken as an initial condition in a Maximum Likelihood estimation, which significantly improves the quality of the final embedding. Apart from its accuracy as an embedding tool, Mercator has the clear advantage of systematically inferring not only node orderings, or angular positions, but also the hidden degrees and global model parameters, and has the ability to embed networks with arbitrary degree distributions. Overall, our results suggest that mixing machine learning and maximum likelihood techniques in a model-dependent framework can boost the meaningful mapping of complex networks.
\end{abstract}

\maketitle

\section{Introduction}

The main hypothesis of network geometry states that the architecture of real complex networks has a geometric origin~\cite{serrano2008self,Krioukov2010,papadopoulos2012popularity}. The nodes of the complex network can be characterized by their positions in an underlying metric space so that the observable network topology---abstracting their patterns of interactions---is then a reflection of distances in this space. This simple idea led to the development of a very general framework able to explain the most ubiquitous topological properties of real networks~\cite{serrano2008self,Krioukov2010}, namely, degree heterogeneity, the small-world property, and high levels of clustering. Network geometry is also able to explain in a very natural way other non-trivial properties, like self-similarity~\cite{serrano2008self} and community structure~\cite{zuev2015emergence,garcia-perez:2018aa,Muscoloni:2018ab}, their navigability properties~\cite{boguna2009navigability,gulyas2015navigable,boguna2010sustaining}, and is the basis for the definition of a renormalization group in complex networks~\cite{GarciaPerez2018}. The geometric approach has also been successfully extended to weighted networks~\cite{allard2017geometric} and multiplexes~\cite{kleineberg2016hidden,Kleineberg2017}.

Beyond being a formal theoretical framework to explain the topology of real networks, network geometry can be used to develop practical applications for real systems, including routing of information in the Internet~\cite{boguna2010sustaining}, community detection~\cite{boguna2010sustaining,serrano2012uncovering,garcia-perez:2016}, prediction of missing links~\cite{papadopoulos2012popularity,WANG2016609,kitsak:2019}, a precise definition of hierarchy in networks~\cite{garcia-perez:2016}, and downscaled network replicas~\cite{GarciaPerez2018}, to name a few.  However, applications require faithful embeddings of real-world networks into the hidden metric space using only the information contained in their topology. Several algorithms have been proposed to solve this problem, most of which either use maximum likelihood estimation techniques~\cite{boguna2010sustaining,papadopoulos2015network_mapping,Papadopoulos:2015fk,blasius2016efficient,Blasius:2018:EES:3209771.3209788}, machine learning~\cite{Alanis-Lobato:2016uq,muscoloni2017machine,Muscoloni:2018aa}, or a combination of both~\cite{Alanis-Lobato2016}.

Maximum likelihood (ML) techniques assume that the network under study has been produced by a given model ---a geometric one--- and finds the value of its parameters that maximize the probability for the model to generate the observed topology. This technique requires finding the coordinates of every nodes in the latent geometry that maximize the likelihood function: a task that, in general, is NP-hard and consequently must rely on heuristics to obtain a reasonable approximate solution. Maximum likelihood methods are therefore generally slow, and their accuracy depends strongly on the chosen heuristic as well as on the quality of the underlying theoretical model.

In contrast, machine learning techniques are fast and model independent, so they can be used to find embeddings of large networks. A promising and accurate method is based on Laplacian Eigenmaps (LE)~\cite{Alanis-Lobato:2016uq,muscoloni2017machine}, originally designed to find dimensional reductions of sets of $n$ points~\cite{Belkin:2001}. The original method requires the definition of Euclidean distances between nodes in $\mathbb{R}^n$, but since no information is available about the ``real Euclidean'' distances between connected pairs of nodes in networks, the use of heuristic arguments is necessary to estimate these distances~\cite{muscoloni2017machine}. A more fundamental problem with machine learning methods is that the embeddings are performed on Euclidean spaces. However, as shown in~\cite{Krioukov2010,papadopoulos2012popularity}, the geometry of real complex networks is better described by hyperbolic rather than Euclidean geometry, where angular coordinates on a circle are a proxy for the similarity between nodes, and their radial coordinates account for their popularity, which is typically measured by their degrees~\cite{serrano2008self}. Machine learning methods are only able to infer the angular coordinates corresponding to the similarity sub-space while radial coordinates have to be inferred using some geometric model. Hence, these methods end up being model dependent as well.

Consequently, both types of methods are very sensitive to the model used to describe the network. References~\cite{papadopoulos2015network_mapping,Papadopoulos:2015fk,Alanis-Lobato:2016uq,muscoloni2017machine,Muscoloni:2018aa} are based on the Popularity$\times$Similarity Optimization (PSO) model described in~\cite{papadopoulos2012popularity}, which uses a simple mechanism to explain the emergence of an effective hyperbolic geometry in growing networks. However, this model can only generate pure power-law degree distributions $P(k) \sim k^{-\gamma}$ with $\gamma>2$, whereas the degree distribution in many real networks shows important deviations from such pure power laws. Moreover, the model does not spontaneously generate the nested hierarchy of self-similar subgraphs with increasing average degree, as observed in real systems~\cite{serrano2008self}.

In this paper, we introduce Mercator, a ready-to-use C++ code \footnote{The code will be available at \texttt{https://github.com/networkgeometry/mercator} upon publication.} that mixes the best of the maximum likelihood and machine learning approaches. The mixing of the two techniques was explored in Ref.~\cite{Alanis-Lobato2016} using the PSO model to maximize the likelihood function. Instead, we use the static version of the same type of Popularity$\times$Similarity geometric models, the $\mathbb{S}^1/\mathbb{H}^2$ model~\cite{serrano2008self,Krioukov2010}, that can accommodate arbitrary degree distributions and can reproduce the self-similarity patterns observed in real networks.  The first step in Mercator is to apply a LE approach, as in Ref.~\cite{muscoloni2017machine}, but using the $\mathbb{S}^1/\mathbb{H}^2$ model instead of the PSO to infer the weights of the Laplacian matrix. Doing so yields a first (and fast) embedding method that already outperforms the one of Ref.~\cite{muscoloni2017machine}. The resulting embedding uses only information about pairs of connected neighbors, and can be further improved by using it as a starting point in a ML optimization---based again on the $\mathbb{S}^1/\mathbb{H}^2$ model---that uses information from both connected and not-connected pairs of nodes. The final result is the most accurate embedding method currently available in the literature. Yet, the final complexity of the method is $\mathcal{O}(N^2)$ for sparse networks with $N$ nodes, which makes it competitive for real applications.

\section{Methodological background}
\subsection{The $\mathbb{S}^1/\mathbb{H}^2$ model}

The $\mathbb{S}^1$ model is the simplest among the class of geometric models~\cite{serrano2008self}. The similarity space is a one dimensional sphere---a circle of radius $R$---where $N$ nodes are distributed with a fixed density, set to one without loss of generality, so that $N=2 \pi R$~\footnote{Notice that in thermodynamic limit the curvature of the circle vanishes and the model is \emph{effectively} defined on $\mathbb{R}^1$.}. Each node is also given a hidden variable $\kappa \in[\kappa_0,\infty)$ proportional to its expected degree. In general, $\kappa$ and the angular position $\theta$ can be correlated and distributed according to an arbitrary distribution $\rho(\kappa,\theta)$. In such case, the model is able to generate community structure~\cite{zuev2015emergence,garcia-perez:2018aa,Muscoloni:2018ab} and can reproduce different degree-degree correlation patterns and clustering spectra.

Once all nodes are assigned a tuple $(\kappa,\theta)$, each pair of nodes is connected with probability
\begin{equation}
  p_{ij}=\frac{1}{1+\left(\frac{d_{ij}}{\mu \kappa_i \kappa_j}\right)^\beta},
  \label{pij}
\end{equation}
where $d_{ij}=R \Delta \theta_{ij}$ is the arc length of the circle between nodes $i$ and $j$ separated by an angular distance $\Delta \theta_{ij}$. Parameters $\mu$ and $\beta$ control the average degree and the clustering coefficient, respectively. The model can be defined using any connection probability as long as it is an integrable function $p(\chi)$ with $\chi=\frac{d}{\mu\kappa \kappa'}$~\cite{serrano2008self}. The particular functional form that we chose here (the Fermi distribution) is the one that defines maximally random ensembles of geometric graphs that are simultaneously clustered, small-world, and with heterogeneous degree distributions.

If nodes are uniformly distributed over the circle, we have $\rho(\kappa,\theta)=\rho(\kappa)/2\pi$. In this case, the choice $\mu = \frac{\beta}{2 \pi \langle k \rangle} \sin \frac{\pi}{\beta}$ guarantees that, in the thermodynamic limit, the expected degree of a node with hidden variable $\kappa$ is $\bar{k}(\kappa)=\kappa$ and the network average degree is $\langle k \rangle=\langle \kappa \rangle$. It is therefore possible to associate unambiguously the hidden variable $\kappa$ with the node degree. For finite systems, however, the values of the hidden variables $\kappa$ must be evaluated numerically. It is also important to notice that the parameter $\mu$ is, in fact, superfluous since it can be absorbed in the definition of $\kappa$; $\kappa$ would then be proportional, but not exactly equal, to the expected degree. As a result, the embedding task only requires the estimation of $2N+1$ parameters: the hidden variables $(\kappa_i,\theta_i)$, $i=1,\cdots,N$, and the parameter $\beta$.

\subsubsection{Hyperbolic representation. The $\mathbb{H}^2$ model}

Quite remarkably, the $\mathbb{S}^1$ model can be expressed as a purely geometric model in the hyperbolic plane. By mapping the expected degree of each node $\kappa_i$ to a radial coordinate as
\begin{equation}
  r_i=\hat{R}-2 \ln{\frac{\kappa_i}{\kappa_0}},
\end{equation}
with $\hat{R} \equiv 2 \ln{\frac{N}{\mu \pi \kappa_0^2}}$, the connection probability becomes
\begin{equation}
  p_{ij}=\frac{1}{1+e^{\frac{\beta}{2}(x_{ij}-\hat{R})}},
\end{equation}
where
\begin{equation}
  x_{ij}=r_i+r_j+2\ln{\frac{\Delta \theta_{ij}}{2}}
\end{equation}
is a good approximation of the hyperbolic distance between two nodes separated by an angular distance $\Delta \theta_{ij}$ and with radial coordinates $r_i$ and $r_j$ \footnote{This approximation is reasonably accurate for pairs of nodes separated by $\Delta \theta_{ij}>\sqrt{e^{-2r_i}+e^{-2r_j}}$, whose fraction converges to one in the thermodynamic limit.}. The connection probability thus becomes a function of the hyperbolic distance alone, which turns the model into a purely geometric one and has important consequences for the global connectivity of the network. For instance, topological shortest paths closely follow geodesic curves in the hyperbolic plane, and can therefore be used to efficiently navigate the network~\cite{boguna2009navigability,boguna2010sustaining}. Furthermore, when the distribution of expected degrees follows a power law of exponent $\gamma$, the radial distribution in the hyperbolic plane is
\begin{equation}
  \rho(r)=\alpha \frac{\sinh{\alpha r}}{\cosh{\alpha \hat{R}}-1}
\end{equation}
with $\gamma=2 \alpha+1$ and $\alpha > 0$. Nodes are therefore homogeneously distributed in the hyperbolic plane for $\gamma=3$ and are quasi-homogeneously distributed for other values of $\gamma$. In this paper, we use the $\mathbb{S}^1$ model for likelihood maximization, and its equivalent $\mathbb{H}^2$ version for visualization purposes.

\subsection{Embedding techniques}

Mercator exploits two different embedding techniques, based on ML and on LE, which are briefly outlined in this section.

\subsubsection{Model-corrected Laplacian Eigenmaps}

Laplacian Eigenmaps was originally designed as a method for dimensional reduction. Given a set of points $\{\mathbf{x}_i \in \mathbb{R}^n, i=1,\cdots,N\}$ with the Euclidean metric, LE finds a mapping of these points $\{\mathbf{x}_i \mapsto \mathbf{y}_i \in \mathbb{R}^m\}$ with $m<n$ such that the loss function
\begin{equation}
  \epsilon=\sum_{i,j} |\mathbf{y}_i-\mathbf{y}_j|^2 \omega(|\mathbf{x}_i-\mathbf{x}_j|^2)
  \label{loss}
\end{equation}
is minimized. Here, $|\mathbf{y}_i-\mathbf{y}_j|$ is the euclidian distance between points $i$ and $j$ in $\mathbb{R}^m$ and $\omega(\cdot)$ is a decreasing function of the distance between the same pair of points in the original Euclidean space $\mathbb{R}^n$. Intuitively, placing pairs of points far apart in $\mathbb{R}^m$ if they were originally close in $\mathbb{R}^n$ increases the loss function Eq.~\eqref{loss}. Minimizing $\epsilon$ should therefore yield a faithful dimensional reduction of the data.

In the case of network embedding, our aim is to find coordinates of nodes in $\mathbb{R}^2$ of a network whose structure can be modeled by the $\mathbb{S}^1$ model. To do so, the weight function $\omega(\cdot)$ is taken to be proportional to the adjacency matrix so that it is only different from zero if nodes $i$ and $j$ are connected. Yet, the weight associated to connected pairs of nodes is still assumed to be a decreasing function of their original Euclidean distance, which must somehow be estimated from the network structure. To do so, we leverage the $\mathbb{S}^1$ model and estimate the expected distance in $\mathbb{R}^2$ (the chord length) of a pair of nodes based on their degrees. The set of coordinates that minimize the loss function $\epsilon$ is the solution of a generalized eigenvalue problem with the Laplacian matrix, for which very fast algorithms exist if the network is sparse~\cite{Lehoucq1996}.

\subsubsection{Maximum likelihood estimation}

Given a real network with adjacency matrix $\{a_{ij}\}$, maximum likelihood estimation finds the values of $\{\kappa_i,\theta_i\}$, $i=1,\cdots,N$, that provide a good match between the $\mathbb{S}^1$ model and the observed network. The posterior probability, or likelihood, that a network specified by its adjacency matrix $\{a_{ij}\}$ is generated by the $\mathbb{S}^1$ model is
%
\begin{equation}
{\cal L}(\{a_{ij}\}| \mathbb{S}^1)=\int \cdots \int {\cal L}(\{a_{ij}\},\{\kappa_i,\theta_i\}| \mathbb{S}^1) \prod_{i=1}^N d\theta_i d \kappa_i,
\end{equation}
%
where the function ${\cal L}(\{a_{ij}\},\{\kappa_i,\theta_i\}| \mathbb{S}^1)$ is the joint probability that the $\mathbb{S}^1$ model generates simultaneously the set of hidden variables $\{\kappa_i,\theta_i\}$ and the adjacency matrix $\{a_{ij}\}$. Using Bayes rule, we then compute the likelihood that the hidden variables $\{\kappa_i,\theta_i\}$ take particular values conditioned on the observed adjacency matrix $\{a_{ij}\}$
%
\begin{align}
{\cal L}(\{\kappa_i,\theta_i\}| \{a_{ij}\},\mathbb{S}^1) & = \frac{{\cal L}(\{a_{ij}\},\{\kappa_i,\theta_i\}|\mathbb{S}^1)}{{\cal L}(\{a_{ij}\}|\mathbb{S}^1)} \nonumber \\
&=\frac{\mbox{Prob}(\{\kappa_i,\theta_i\}){\cal L}(\{a_{ij}\}|\{\kappa_i,\theta_i\},\mathbb{S}^1)}{{\cal L}(\{a_{ij}\}|\mathbb{S}^1)},
\label{likelihood}
\end{align}
%
where
\begin{equation}
\mbox{Prob}(\{\kappa_i,\theta_i\})=\prod_{i=1}^N \rho(\theta_i,\kappa_i)
\end{equation}
is the prior probability density function of the hidden variables,
\begin{equation}
{\cal L}(\{a_{ij}\}|\{\kappa_i, \theta_i\}, \mathbb{S}^1)=\prod_{i<j} p_{ij}^{a_{ij}} \left(1-p_{ij}\right)^{1-a_{ij}},
\label{likelihood2}
\end{equation}
is the probability that the $\mathbb{S}^1$ model generates the adjacency matrix $\{a_{ij}\}$ conditioned on the hidden variables $\{\kappa_i, \theta_i\}$, and $p_{ij}$ is the connection probability given by Eq.~\eqref{pij}.

If we have information about the prior distribution of hidden variables, $\mbox{Prob}(\{\kappa_i,\theta_i\})$, Bayesian estimators can be obtained by maximizing the likelihood in Eq.~(\ref{likelihood}). However, in most cases, prior information is not available. We then use an improper prior distribution $\mbox{Prob}(\{\kappa_i,\theta_i\})=cte$, and obtain the maximum likelihood estimator as the set of values $\{\kappa_i^*, \theta_i^*\}$ that maximize Eq.~\eqref{likelihood2} or, equivalently, its logarithm
\begin{align}
  \ln{\cal L}(\{a_{ij}\}&|\{\kappa_i, \theta_i\}, \mathbb{S}^1) \nonumber \\
  & =\sum_{i<j} \left[a_{ij} \ln{p_{ij}}+(1-a_{ij}) \ln{\left(1-p_{ij}\right)} \right] \ .
  \label{likelihoodfinal}
\end{align}
The maximization with respect to the expected degrees $\kappa$ can be performed semi-analytically. The derivative of Eq.~(\ref{likelihoodfinal}) with respect to the expected degree $\kappa_l$ of node $l$ is
\begin{equation}
  \frac{\partial}{\partial \kappa_l} \ln{\cal L}(\{a_{ij}\}|\{\kappa_i, \theta_i\}, \mathbb{S}^1)=\frac{\beta}{\kappa_l}\sum_{i \ne l} (a_{il}-p_{il}) \ ,
\end{equation}
%
where the second term on the right hand side is the expected degree of node $l$, and the first term is its actual degree $k_l$. The value $\kappa_l^*$ that maximizes the likelihood is therefore the solution of
\begin{equation}
  k_l=\sum_{i \ne l} p_{il} \ .
  \label{kappasML}
\end{equation}
The term on the right hand side can be evaluated in the model assuming a homogeneous angular distribution of nodes on the circle. We use this method to provide estimates of the expected degrees that are then used to maximize the likelihood function with respect to the angular coordinates, as explained in Sec.~\ref{maxlike}.

\section{Mercator at a glance}

We have now all the theoretical background to briefly describe Mercator; the full details are given at Secs.~\ref{sec:sketch}--\ref{sec:adjustment_hidden_degrees}. Given a network with adjacency matrix $\{a_{ij}\}$, we first measure its average degree $\langle k \rangle$, average clustering coefficient $\bar{c}$, and all individual nodes' degrees $\{ k_i \;, i=1,\cdots,N\}$. Second, we estimate hidden degrees and parameters $\beta$ and $\mu$. Third, we estimate the angular ordering of nodes using the model-corrected LE, and adjust the angles according to the expected angular separation between consecutive nodes given by the $\mathbb{S}^1$ model. This yields Mercator's fast mode version, which produces a first embedding. Fourth, the angular coordinates are refined using ML. Finally, hidden degrees are readjusted given the newly inferred angular positions. All the steps together conform Mercator refined mode. More precisely, Mercator executes the following steps.

\subsection{Fast mode}
\begin{enumerate}
  \item Propose an approximate value for $\beta$ and compute $\mu = \frac{\beta}{2 \pi \langle k \rangle} \sin \frac{\pi}{\beta}$.
  \item Using Eqs.~\eqref{pij}~and~\eqref{kappasML}, adjust the hidden variables $\{\kappa_i\}$ such that the expected degree of each node in the $\mathbb{S}^1$ model matches the observed degree in the original network. This step assumes that nodes are homogeneously distributed and uses the values of $\beta$ and $\mu$ from step 1. The initial guess is $\kappa_i=k_i$ (the degree of node $i$ in the original network).
  \item Using results from steps 1 and 2, evaluate the theoretical value of the average clustering coefficient of the network in the $\mathbb{S}^1$ model. If this value differs from the one measured for the original network, adjust the value of $\beta$ and return to step 1. Otherwise, proceed to step 4.
  \item For every connected pair of nodes with hidden variables $\kappa_i$ and $\kappa_j$ and original degrees $k_i,k_j>1$, estimate their expected chord length in $\mathbb{R}^2$ as $d_{ij}=2\sin{\frac{\langle \Delta \theta_{ij}\rangle }{2}}$, where $\langle \Delta \theta_{ij}\rangle$ is the expected angular separation between connected nodes $i$ and $j$ in the $\mathbb{S}^1$ model.
  \item Construct a weighted Laplacian matrix $L_{ij}=D_{ij}-\omega_{ij}$, where $D$ is the diagonal matrix with entries $D_{ii}= \sum_{j} \omega_{ij}$ and weights are given by $\omega_{ij}=a_{ij} e^{-d_{ij}^2/t}$ with $t$ being the variance of $d_{ij}$. Then solve the generalized eigenvalue problem
  \[
    L\mathbf{v}=\lambda D \mathbf{v}\ .
  \]
  We note $\mathbf{v_1}=(v_{1,1},v_{1,2},\cdots,v_{1,N})$ and $\mathbf{v_2}=(v_{2,1},v_{2,2},\cdots,v_{2,N})$ the first two eigenvectors with the smallest nonzero eigenvalues.
  \item Assign an angular position to each node $i$ as $\theta_i = \mathrm{atan2}(v_{2,i},v_{1,i})$.
  \item Make a sorted list of the nodes based on their angular position $\{\theta_i\}$. Nodes of degree 1 that were excluded at step 4 are now reinserted in the sorted list randomly before or after their unique neighbor. Note that the angular coordinates computed at step 6 are only used to determine the order in which nodes are located angularly. Their precise angular coordinates are evaluated at the next step.
  \item For each pair of consecutive nodes in the list, evaluate its expected angular separation using the $\mathbb{S}^1$ model conditioned on whether the two nodes are connected or not, their hidden variables $\kappa_i$ and $\kappa_j$, and the fact that they are consecutive. Finally, all gaps are normalized to sum up to $2\pi$. This produces a set of angular coordinates for each node $\{\theta_i \; , i=1,\cdots, N\}$.
\end{enumerate}
These 8 steps summarize a fast and accurate embedding procedure that, as discussed in Sec.~\ref{sec:validation_synthetic_networks}, already outperforms current state-of-the-art methods. The next section explains how its accuracy can be further increased.

\begin{figure}
  \begin{center}
    \includegraphics[width=\columnwidth]{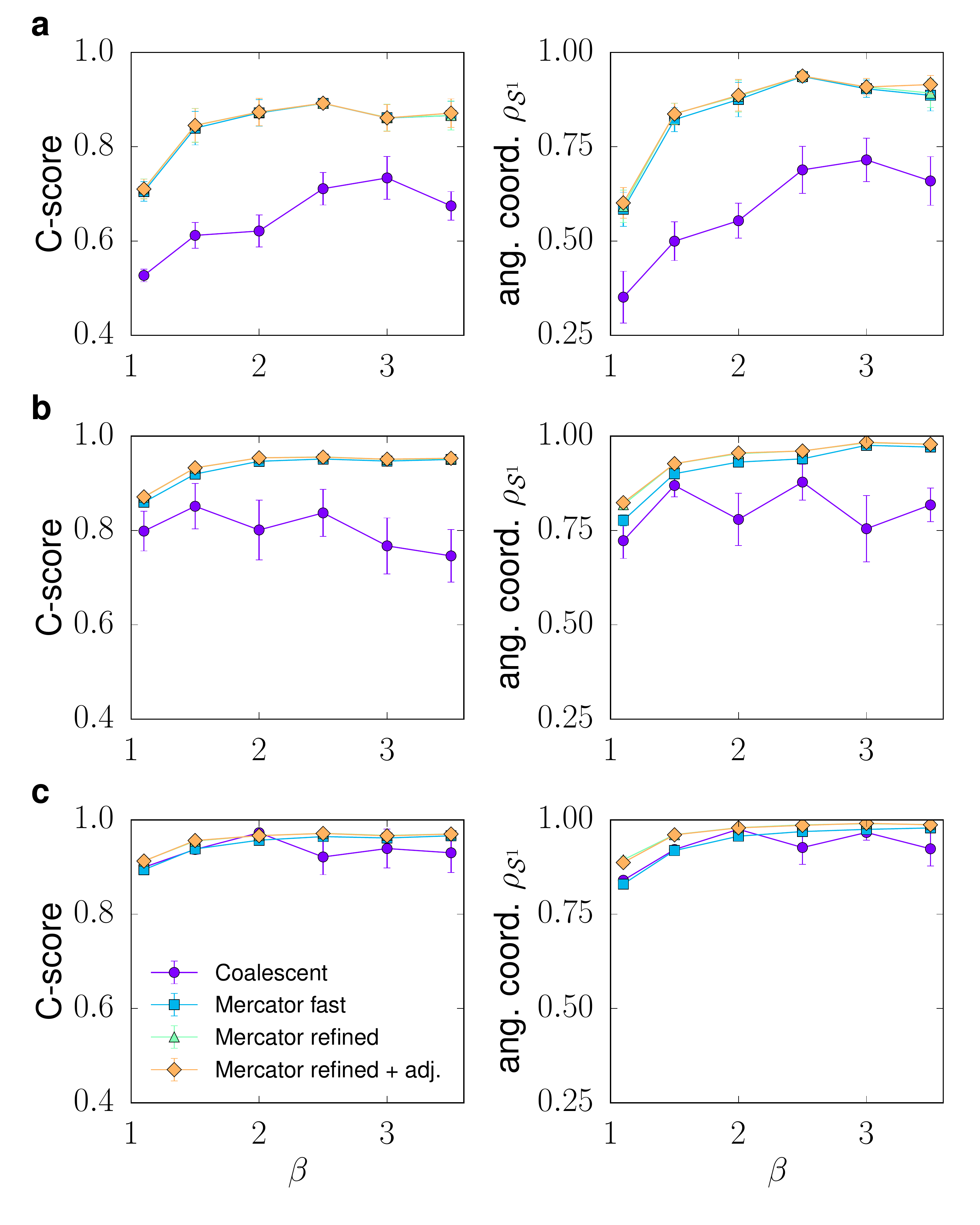}
    \caption{{Comparison between the angular coordinates inferred by different embedding methods in terms of the C-score (left) and the the Pearson correlation coefficient (right) between original and inferred angular coordinates as a function of the clustering parameter $\beta$. Every row corresponds to a different value of $\langle k \rangle$: \textbf{a.} 4, \textbf{b.} 8, and \textbf{c.} 12. For every value of the parameters, we generated and embedded 10 synthetic networks of size $N = 1000$ and with power-law degree distribution exponent $\gamma = 2.5$. The plots show the resulting averages and standard deviations.}
    \label{fig:Cscore-S1Pearson}}
  \end{center}
\end{figure}

\subsection{Refined mode}

The embedding can be significantly improved with ML techniques using the embedding obtained with the fast mode of Mercator as initial conditions. This is due to the fact that ML uses the information contained both in the presence \textit{and} absence of links in the network, whereas LE only relies on the information conveyed by the presence of links.  The major drawback of ML is the complex configuration space that needs to be explored to find the optimal embedding. However, if the starting point of the exploration is good enough, the maximization of the likelihood function is easy and efficient. Thus, starting from the embedding obtained with the fast mode, we proceed as follows.
\begin{enumerate}
  \setcounter{enumi}{8}
  \item Extract the onion decomposition of the network~\cite{Hebert-Dufresne:2016aa} and sort nodes accordingly, starting with the node in the deepest layer~\footnote{The onion decomposition is a generalization of the $k$-core decomposition that provides the \emph{internal organization} of each $k$-shell. It therefore offers a more precise way to order nodes than based on their position in the $k$-core decomposition alone.}. Doing so allows the likelihood optimization phase to begin with the most central nodes (based on mesoscale topological information) thereby greatly facilitating the finding of an acceptable local maximum in the configuration space.
  \item For each node in the sorted list, find the average angular coordinate of its neighbors.
  \item Sample $\mathcal{O}(\ln{N})$ angular positions around this average value using a normal distribution whose standard deviation is set to half of the angular distance between this average value and the farthest neighbor.
  \item Compute the local log-likelihood of the sampled angular positions
  \begin{equation}
  \ln \mathcal{L}_i = \sum \limits_{j \neq i} a_{ij} \ln p_{ij} + (1 - a_{ij} ) \ln (1 - p_{ij})\ ,
  \end{equation}
  where $p_{ij}$ is computed with Eq.~\eqref{pij}, and set the new angular position of the node to the sampled angle with highest log-likelihood.
  \item Once the position of every node has been optimized once, repeat step 2 to find a better estimate of the hidden variables $\kappa_i$ using the newly inferred angular positions. This last step is optional, although it generally leads to substantial improvements of the final embedding.
\end{enumerate}

\begin{figure}
  \begin{center}
    \includegraphics[width=\columnwidth]{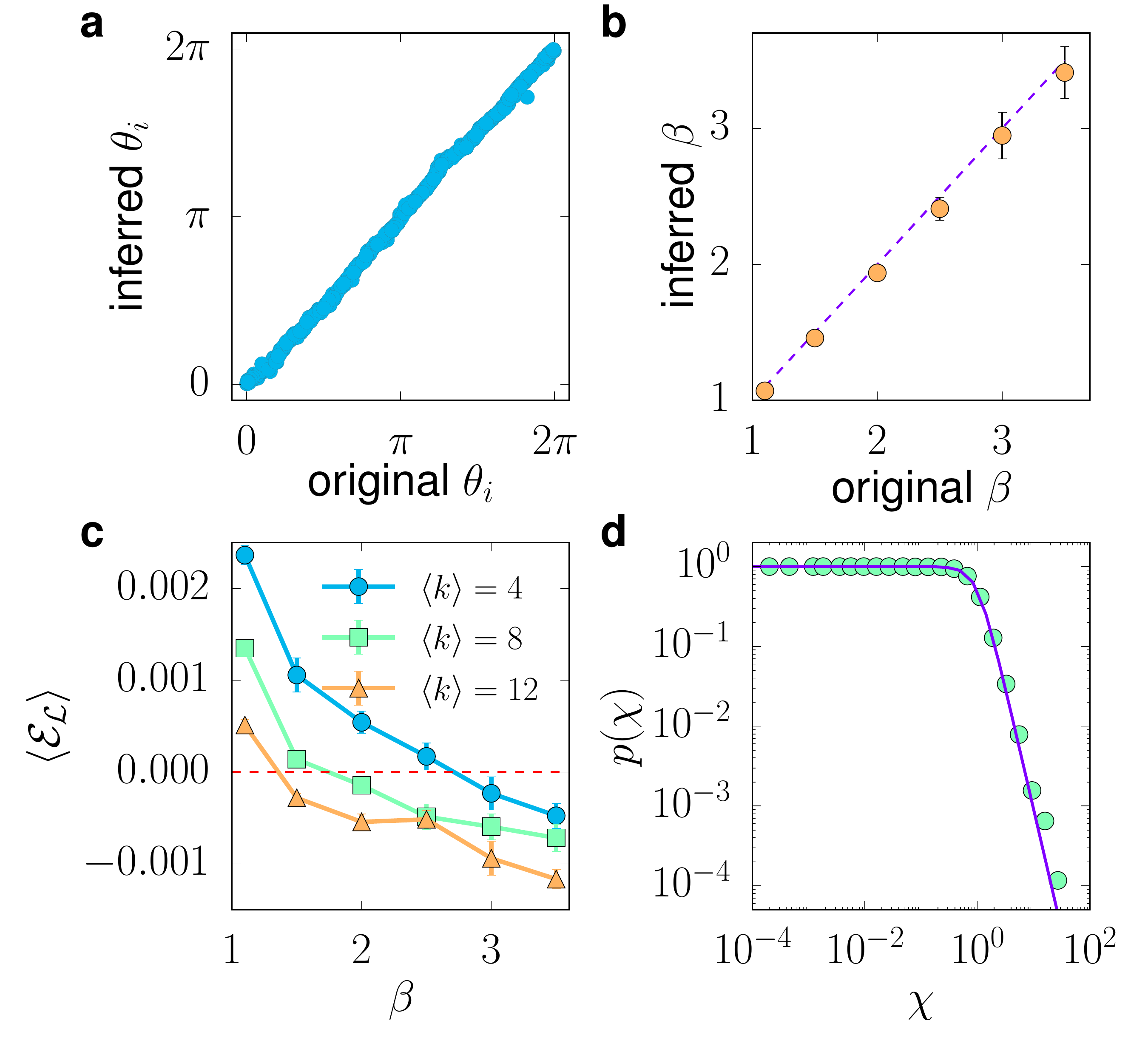}
    \caption{\textbf{a.} Example of the inferred angular coordinates vs. original angular coordinates for a network with $\langle k \rangle = 12$ and $\beta = 3$.
    \textbf{b.} Inferred vs. original values of the global parameter $\beta$ obtained from the embeddings of the networks in Fig.~\ref{fig:Cscore-S1Pearson}. \textbf{c.} Relative likelihood difference $\langle \mathcal{E}_{\mathcal{L}} \rangle = \langle \bar{\mathcal{L}}_{\mathrm{infer}} / \bar{\mathcal{L}}_{\mathrm{real}} \rangle -1$ averaged over all the networks of a given $\langle k \rangle$ and $\beta$ embedded with Mercator (refined mode with final adjustment of the $\{\kappa_i\}$ at step \#13). In many cases, Mercator is able to find embeddings with likelihoods above those that generated the networks. In both plots, the errorbars represent the corresponding standard deviations. \textbf{d.} Empirical connection probability (circles) compared to the theoretical curve for the network in \textbf{a}.
    \label{fig:Model_inference}}
  \end{center}
\end{figure}

\section{Validation in synthetic networks} \label{sec:validation_synthetic_networks}

We test Mercator using synthetic networks of different average degrees and clustering coefficients generated with the $\mathbb{S}^1$ model, and consider several quality measures to check the accuracy of the embeddings. We first focus on its capability to recover the angular coordinates. Figure~\ref{fig:Cscore-S1Pearson} shows the C-score~\cite{muscoloni2017machine}, defined as the fraction of pairs of nodes that are correctly ordered in the circle, as well as the Pearson correlation coefficient between the real and inferred angular coordinates~\footnote{Notice that the model is invariant with respect to global rotations and inversions of the angular coordinates. Therefore, we consider the maximal Pearson correlation coefficient over all such possible transformations}. The results of the two versions of Mercator are compared with those obtained using the Coalescent embedding presented in Ref.~\cite{muscoloni2017machine}, which was reported to give the best node orderings with respect to other embedding algorithms in the literature. Notice that Mercator is able to outperform the results even in its fast mode, especially for networks with a low average degree. As an example, Fig.~\ref{fig:Model_inference}\textbf{a} depicts the inferred angular coordinates versus the real ones for one of the networks considered.

\begin{figure*}
  \begin{center}
    \includegraphics[width=\linewidth]{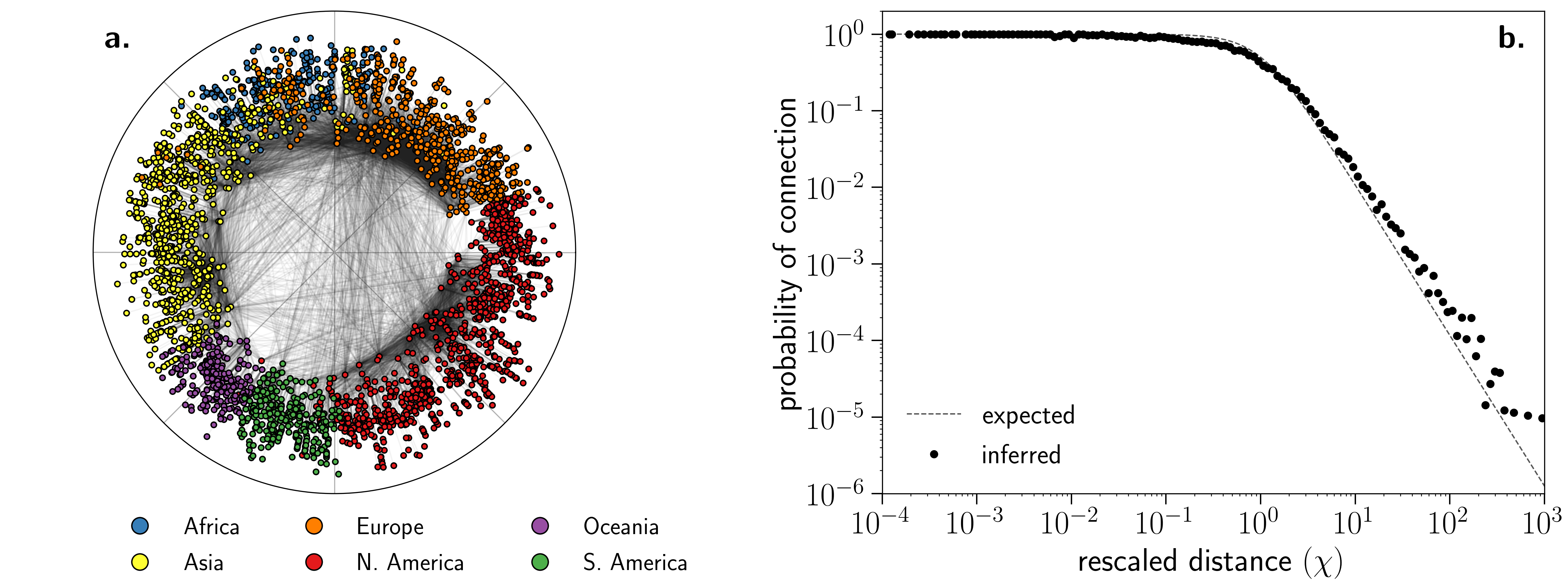}
    \caption{\textbf{a.} Hyperbolic embedding of the world airports network obtained by Mercator in the refined mode. Nodes are colored according to the continent in which they are located, information that is not used to obtain the embedding. \textbf{b.} Comparison of the expected connection probability based on the inferred value of $\beta$ (expected) and the actual connection probability computed with the inferred hidden variables $\{k_i, \theta_i\}$ (inferred).
    \label{fig:airport:pij}}
  \end{center}
\end{figure*}

Mercator also has the clear advantage of systematically inferring the hidden degrees and global model parameters. In Fig.~\ref{fig:Model_inference}\textbf{b} we show that the inference of $\beta$ is very precise for all the synthetic networks considered in this section. This has important implications for applications that require finding a good congruency between the network and the model. Indeed, Mercator is able to find embeddings with very high likelihoods. To quantify this, we consider the relative likelihood difference $\mathcal{E}_{\mathcal{L}} = \bar{\mathcal{L}}_{\mathrm{infer}} / \bar{\mathcal{L}}_{\mathrm{real}} -1$, where $\bar{\mathcal{L}} \equiv \mathcal{L}^{2/N(N-1)}$ is the geometric average of the likelihood over all pairs of nodes. Hence, a positive (negative) $\mathcal{E}_{\mathcal{L}}$ indicates that the inferred embedding has a higher (lower) likelihood than the real coordinates and model parameters. Strikingly, for low values of $\beta$, the embeddings found by Mercator have $\mathcal{E}_{\mathcal{L}} > 0$ (Fig.~\ref{fig:Model_inference}\textbf{c}). Finally, Fig.~\ref{fig:Model_inference}\textbf{d} presents an example of the empirical connection probability (fraction of connected pairs as a function of the rescaled distance $\chi = d/(\mu \kappa \kappa')$), which is extremely congruent with Eq.~\eqref{pij}. Put together, these results reveal that Mercator is not just the most accurate algorithm in terms of the reconstruction of the angular coordinates of synthetic networks---arguably the most difficult aspect of the embedding problem---, but it also determines correctly all other model parameters, including hidden degrees.

\section{Embedding of real networks}
\label{sec:real_networks}

Another strength of Mercator is its ability to embed networks with arbitrary degree distributions. As an illustration, we embedded several real-world complex networks from different domains whose degree distributions include clean scale-free, heavy-tailed, and arbitrary distributions. More specifically, the networks under study are: the world airport network~\footnote{Downloaded from \texttt{https://openflights.org/data.html}.}, the neural network of the visual cortex of the Drosophila Melanogaster at the neuron level~\cite{Takemura2013}, the neural network of the C. Elegans worm~\cite{Varshney2011}, a human connectome~\cite{avena2018communication,Hagmann2008}, the metabolic network of the bacterium E. Coli~\cite{Orth2011,serrano2012uncovering}, the world trade web~\cite{garcia-perez:2016}, a US commute network~\cite{Grady2012}, a cargo ships network~\cite{Kaluza2010}, a US commodities network~\cite{Grady2012}, and the Internet at the Autonomous Systems level~\cite{boguna2010sustaining}.

One particularly telling example is the airports network whose truncated power-law degree distribution with exponent $\gamma<2$ cannot be easily embedded with methods based on the PSO model. In the case of real networks, we do not have access to the ``real'' coordinates to compare with those obtained from our embeddings. Yet, in some cases, metadata related to the similarity between nodes is available and can be used to test whether an embedding is meaningful or, instead, is an artifact of the algorithm. In the case of the airports network, geography is such metadata. Figure~\ref{fig:airport:pij}\textbf{a} shows the hyperbolic embedding obtained by Mercator in the slow mode with nodes colored according to the continent they belong to (separating North and South America). Airports belonging to the same continent appear clustered in similar angular positions, thus supporting the relation between the angular space of the embedding and similarity among nodes. Similar analyses were carried out for the Internet at the autonomous systems level~\cite{boguna2010sustaining}, metabolic networks~\cite{serrano2012uncovering}, and the world trade web~\cite{garcia-perez:2016}. A strong correlation between the angular distribution of points and available metadata was found in all cases. In light of these results, we conclude that our geometric embeddings are meaningful and capture attributes that contribute to the similarity among the elements of complex networks.

\begin{figure*}
  \begin{center}
    \includegraphics[width=\linewidth]{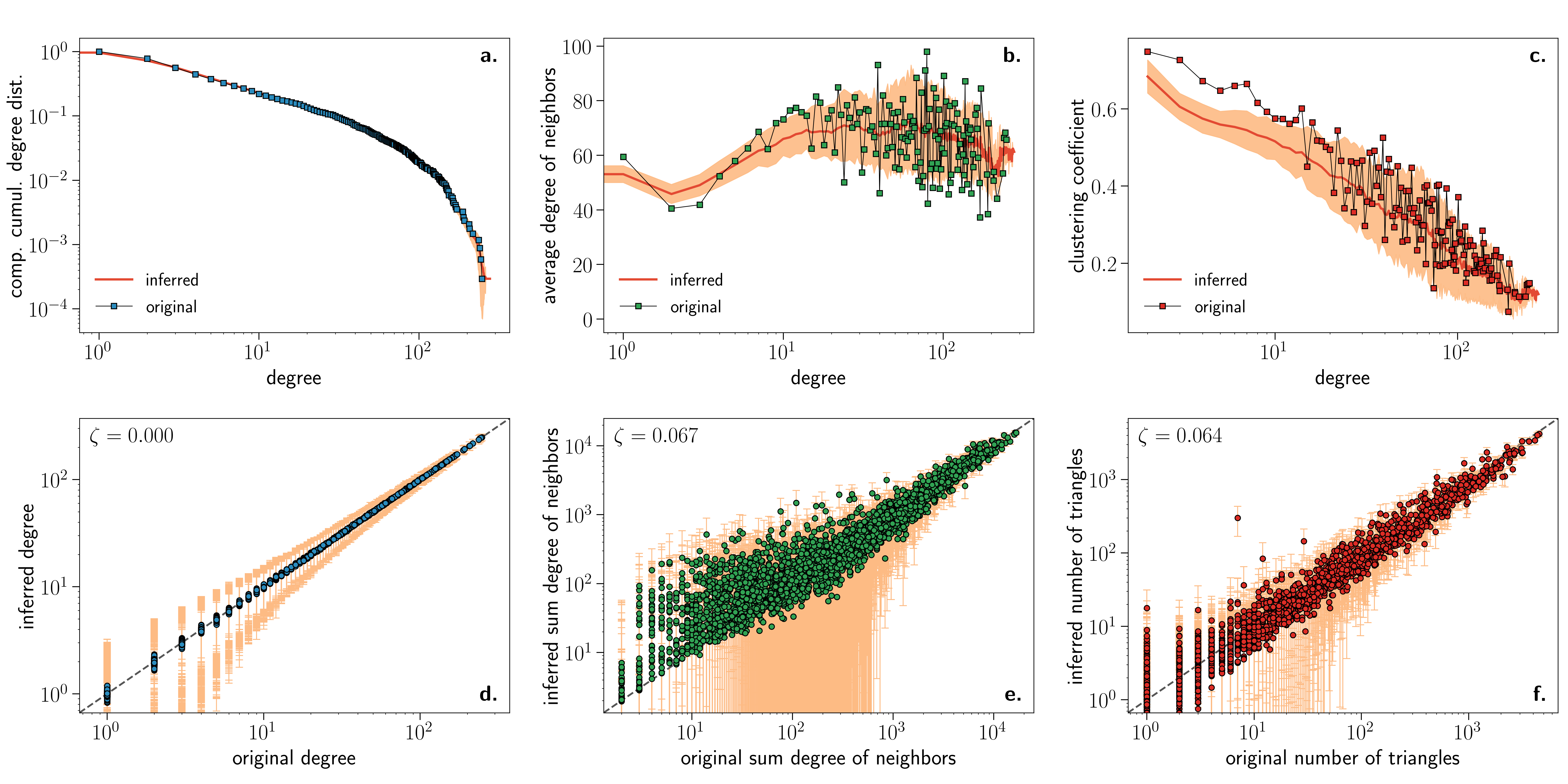}
    \caption{Topological validation of the embedding of the airports network. The first row shows \textbf{a.} the complementary cumulative degree distribution, \textbf{b.} the average nearest neighbors degree, $\bar{k}_{nn}(k)$, and \textbf{c.} the clustering spectrum, $\bar{c}(k)$. Symbols correspond to the value of these quantities in the original network, whereas the red lines indicate an estimate of their expected values in the ensemble of random networks inferred by Mercator. This ensemble was sampled by generating 100 synthetic networks with the $\mathbb{S}^1$ model and the inferred parameters and positions by Mercator. The orange regions correspond to an estimate of the $2\sigma$ confidence interval around the expected values.
    The second row shows scattered plots of \textbf{d.} the degree of every nodes, \textbf{e.} the sum of the degrees of their neighbors, and \textbf{f.} the number of triangles to which they participate. The plots show the estimated values of these three measures in the same ensemble of random networks considered above versus the corresponding values in the original network. The error bars show the $2\sigma$ confidence interval around the expected values. The quantity $\zeta$ corresponds to the fraction of nodes for which the value measured on the original network lies outside the $2\sigma$ confidence interval.
    \label{fig:airport}}
  \end{center}
\end{figure*}

Beyond this qualitative agreement, we tested the extent to which the embedding inferred by Mercator is accurate enough to reproduce the topology of the airports network. To do so, we first compare the expected connection probability Eq.~\eqref{pij} with the inferred value of $\beta$ against the empirical connection probability, computed using the inferred coordinates of the nodes ($\{\kappa_i, \theta_i\}$). This remarkable agreement confirms that the rescaled distance $\chi$ provides a meaningful measure to characterize the interaction between nodes in the network. Second, we used the set of coordinates $\{ \kappa_i,\theta_i \}$ as well as the parameters $\beta$ and $\mu$ to generate an ensemble of synthetic networks using Eq.~\eqref{pij}. We then compared several topological properties of this ensemble with those measured on the original network. Specifically, the first row of Fig.~\ref{fig:airport} shows the results for the complementary cumulative degree distribution $P_c(k)$, the average nearest neighbors' degree $\bar{k}_{nn}(k)$, and the clustering spectrum $\bar{c}(k)$. The final adjustment of hidden degrees in step \#13 of the Mercator algorithm strongly enhances the reproduction of the degree distribution. Notice that the $\mathbb{S}^1$ model does not include any mechanism to control degree-degree correlations or the shape of the clustering spectrum (recall that $\beta$ is chosen based on the average clustering coefficient only). Yet, the generated network ensemble reproduces these two quantities with remarkable precision. This is particularly interesting in the case of the average nearest neighbors' degree, which shows a non-trivial assortative behavior for low degrees both in the real network and the ensemble. This suggests that the non-uniform angular distribution of nodes inferred by Mercator (and so the network geometric properties) is partly responsible for the observed degree-degree correlations in real complex networks.

The ensemble of synthetic networks generated from the estimated geometric parameters of the airports network performs also very well at reproducing topological properties of individual nodes. The second row of Fig.~\ref{fig:airport} shows scattered plots of the degree of a given node, the sum of degrees of its neighbors, and number of triangles attached to it in the generated network ensemble versus the same quantities measured on the original network. For each node, we also compute the $2\sigma$ confidence interval, and $\zeta$ shows the fraction of nodes whose original property (degree, sum of neighbors' degree, number of triangles) lies outside this interval. Results show that the fraction of points outside this interval is around $6\%$, which is consistent with the $2\sigma$ (or $\approx 95\%$) confidence interval. These results, supported by those presented in the Supplementary Information, clearly illustrate the accuracy of the embeddings provided by Mercator. To the best of our knowledge, such accuracy cannot be obtained with other existing embedding methods.

\section{Computational complexity}

\begin{figure}
  \begin{center}
    \includegraphics[width=\columnwidth]{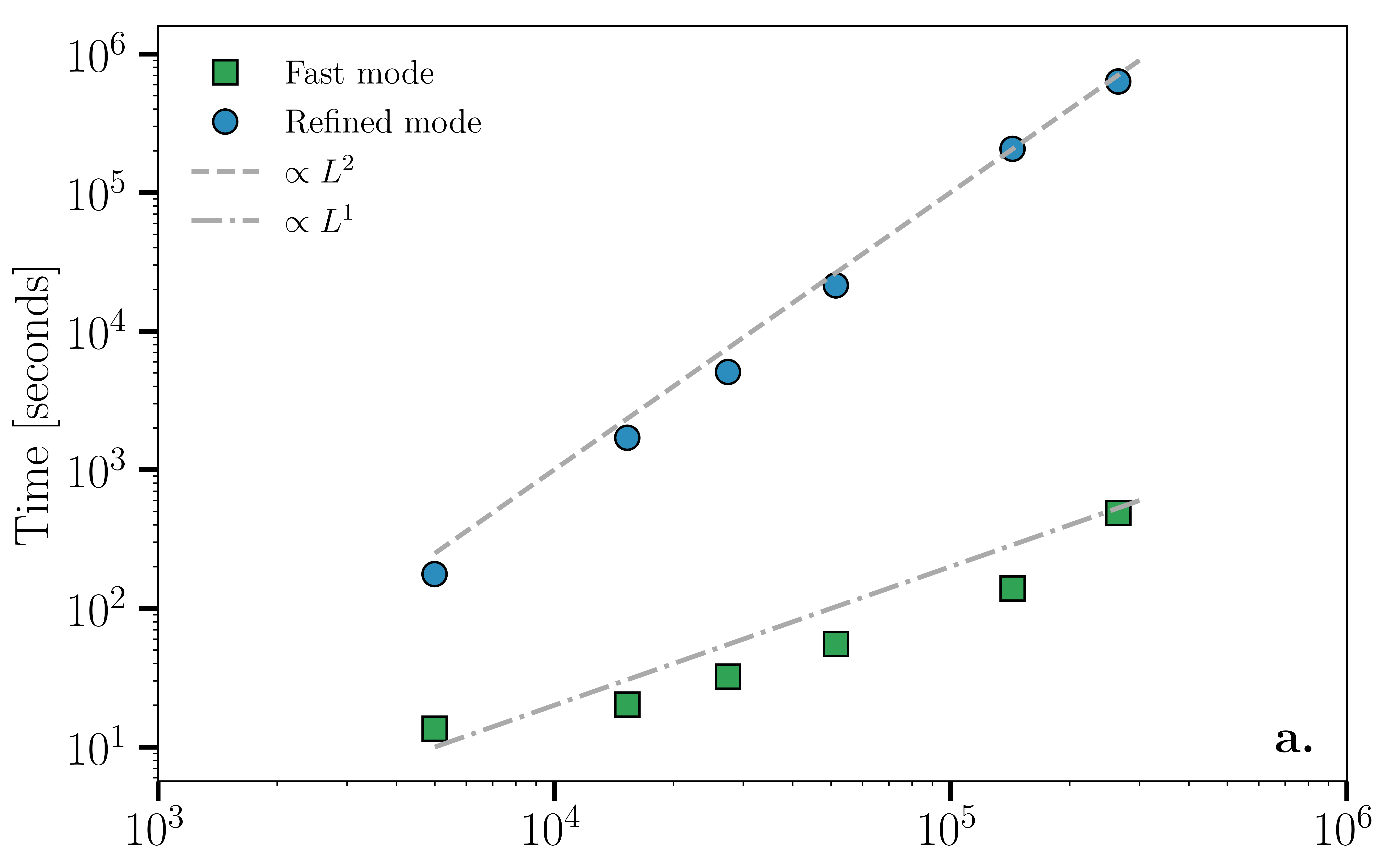}
    \includegraphics[width=\columnwidth]{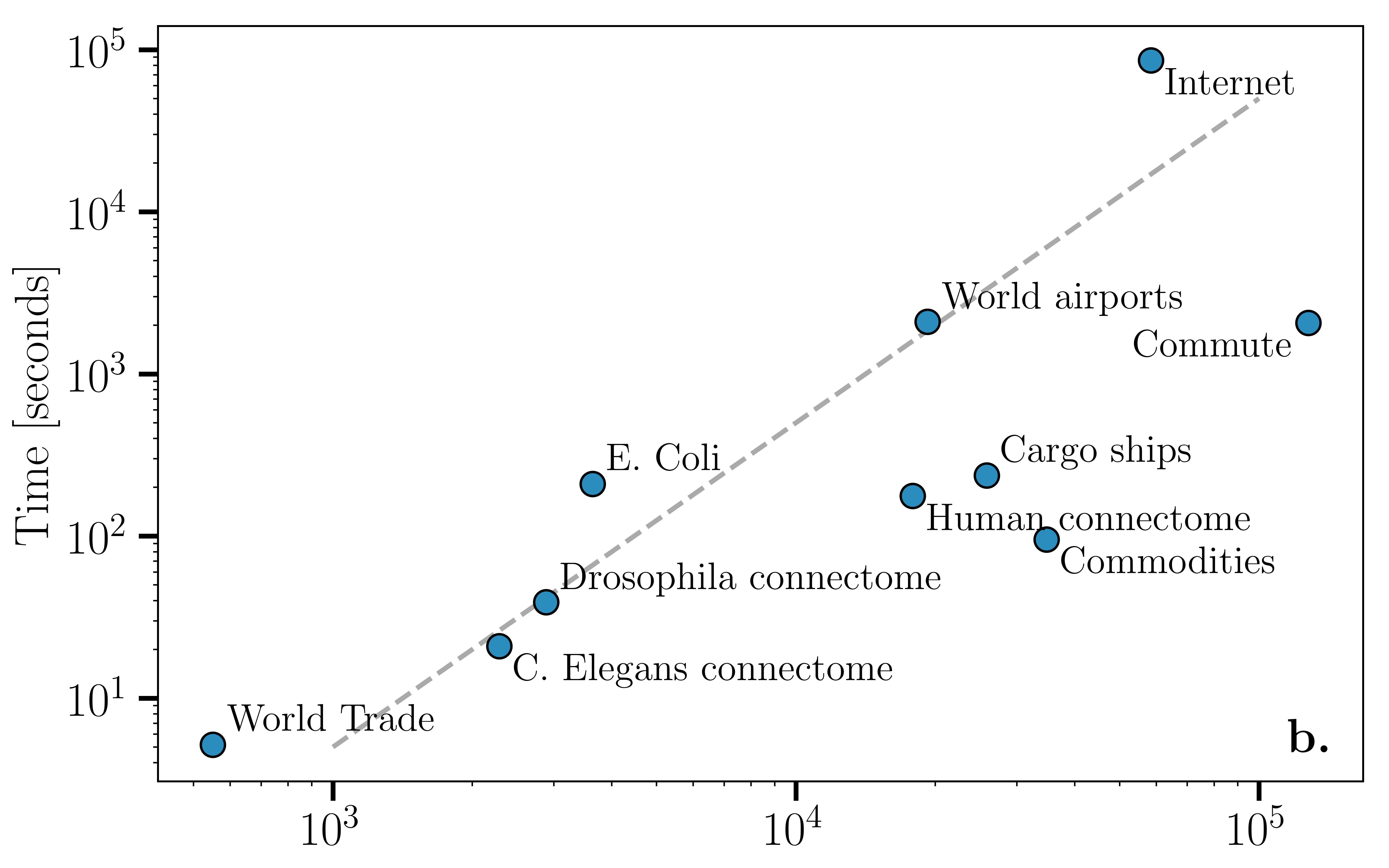}
    \includegraphics[width=\columnwidth]{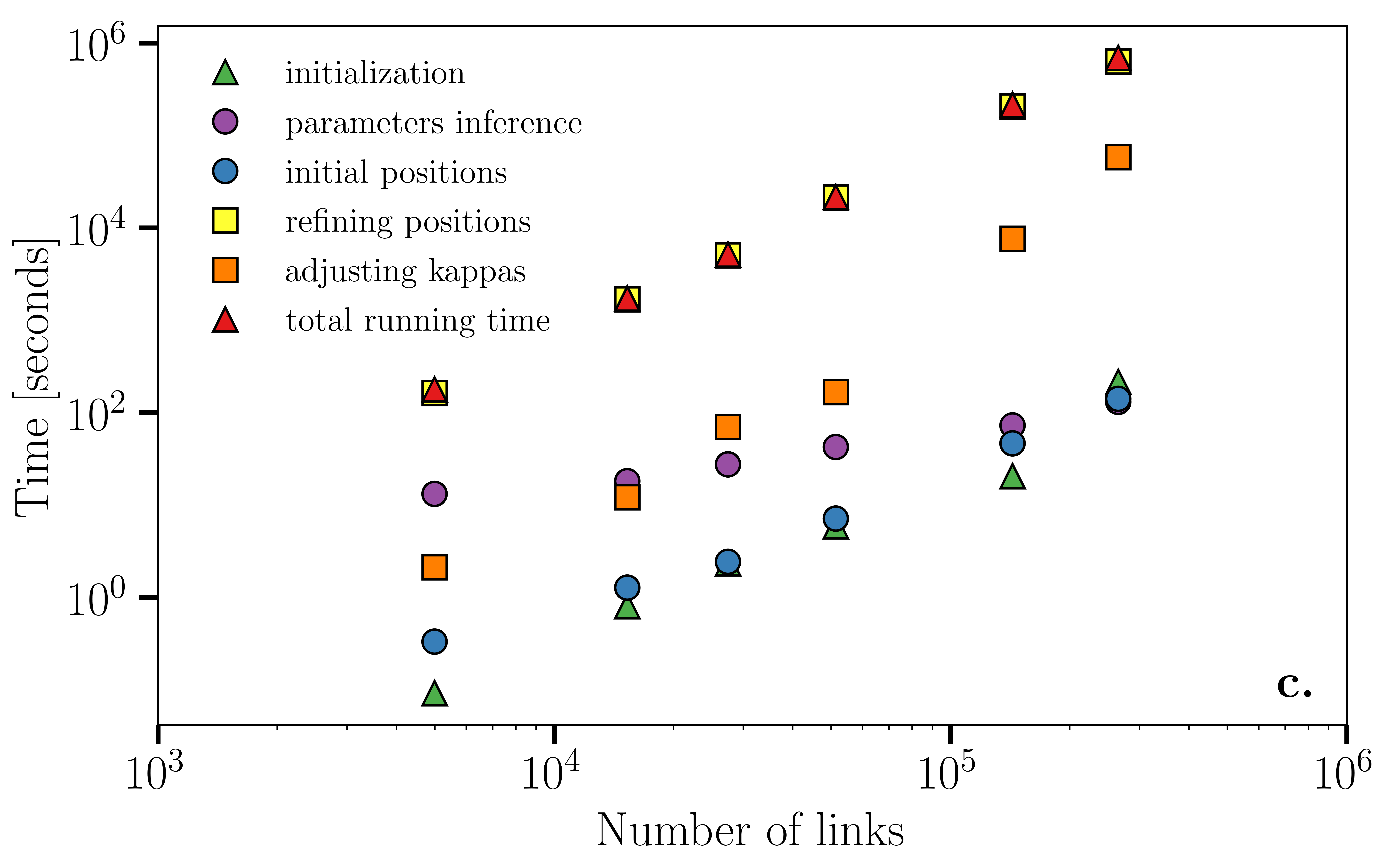}
    \caption{Comparison of the computational complexity of Mercator expressed in terms of the running time versus the number of links in the network ($L$).
    \textbf{a.} Comparison between Mercator's fast and refined modes using synthetic networks generated with the $\mathbb{S}^1$ model with a power law distribution for the expected degrees $\kappa$ with exponent $\gamma=2.2$, $\beta=2$ (clustering) and $\langle k \rangle=10$.  Dashed lines proportional to $L^\alpha$ were added to guide the eye. Higher values of $\gamma$ lead to smaller running times but similar scaling behavior as the number of links increases.
    \textbf{b.} Computational complexity of Mercator's slow mode applied to the real network datasets presented in Sec.~\ref{sec:real_networks}. A line proportional to $L^2$ has been added to guide the eye.
    \textbf{c.} Mercator's computational complexity broken down into each each step of the algorithm. The same synthetic networks as in \textbf{a} were used.
    \label{fig:time_complexity}}
      \end{center}
\end{figure}

We now support our claim that the computational complexity of Mercator scales as $\mathcal{O}(N^2)$ for sparse networks composed of $N$ nodes (and $L=\langle k \rangle N / 2$ links). Figure~\ref{fig:time_complexity}\textbf{a} and \ref{fig:time_complexity}\textbf{b} show the running time in seconds in function of the number of links ($L$) for both synthetic and real networks. In both cases, we find that Mercator's refined mode does indeed roughly scale as $L^2$ although it is clear that other topological properties influence the final total running time. With respect to the fast mode, we find that the computational complexity is roughly linear within the range in the number of links that was considered. Finally, Fig.~\ref{fig:time_complexity}\textbf{c} breaks down the running time into the time spent in each of the Mercator major steps, and doing so shows how most of the running time is spent during the ML step.

\section{Conclusions}

In this work, we introduce and deliver the full code of Mercator, the most accurate method to embed complex networks into their latent metric spaces. We showed that the quality of the embeddings can be significantly improved by a proper combination of machine learning techniques and powerful statistical methods. Thanks to this combination, Mercator is able to overcome some of the drawbacks of other techniques which, for instance, require perfect power laws in the whole domain of degrees, a condition that is not met by many real networks. Our results also indicate that the obtained embeddings are able to recover ground truth information not contained in the network topology. We expect Mercator to become a standard tool within the toolbox of network scientists and anybody interested in retrieving information from big data systems admitting a network representation.

\begin{acknowledgments}
  We acknowledge support from the project {\it Mapping Big Data Systems: embedding large complex networks in low-dimensional hidden metric spaces} -- Ayudas Fundaci\'on BBVA a Equipos de Investigaci\'on Cient\'{\i}fica 2017; James S. McDonnell Foundation Scholar Award in Complex Systems; the ICREA Academia prize, funded by the Generalitat de Catalunya; Ministerio de Econom\'{\i}a y Competitividad of Spain, project no. FIS2016-76830-C2-2-P (AEI/FEDER, UE). GGP acknowledges financial support from the Academy of Finland via the Centre of Excellence program (Project no.~312058 as well as Project no.~287750), and from the emmy.network foundation under the aegis of the Fondation de Luxembourg.  AA acknowledges financial support from the project Sentinelle Nord of the Canada First Research Excellence Fund, from ``la Caixa'' Foundation and from the Spanish ``Juan de la Cierva-incorporaci\'on'' program (IJCI-2016-30193).
\end{acknowledgments}

\appendix
\section{Mercator in details}
We now provide the full details of Mercator.

\label{MercatorFull}
\subsection{Sketch of the method}\label{sec:sketch}
The $\mathbb{S}^{1}$ model has two global parameters that need to be inferred; $\mu$, controlling the average degree and $\beta$, which determines the level of clustering in the network. In addition, every node $i$ is assigned two hidden variables: a hidden degree $\kappa_{i}$ and an angular coordinate $\theta_{i}$. The following method finds the values of $\mu$, $\beta$ and $\kappa_{i}$ for which the expected degrees in the model $\bar{k}(\kappa_i)$, that is, in synthetic networks generated with uniformly distributed angular positions, equal the observed degrees in the real network and, moreover, the expected mean local clustering of the embedding matches the real value. To that end, some values of $\mu$ and $\beta$ are proposed. Next, the corresponding $\kappa_{i}$ are calculated. Finally, the expected clustering coefficient is computed and $\beta$ is adjusted if the predicted value is not within an acceptable range of the original value.

The method relies on the assumption that all nodes with the same degree have the same hidden degree. Therefore, the first preliminary step is reading the network and counting the number of nodes in every degree class $k$, that we denote by $N_k$.

\subsection{Inferring the hidden degrees}\label{sec:inf_hid_degs}
This step assumes some given value of $\beta$ and the corresponding $\mu = \frac{\beta}{2 \pi \langle k\rangle} \sin \frac{\pi}{\beta}$, where $\langle k \rangle$ is the observed average degree. We then assign to every degree class the hidden degree given by $\kappa(k) = k$ as the initial guess. The aim of the following algorithm is to adjust this relation so that $\bar{k}(\kappa(k)) = k \pm \epsilon$, where $\epsilon$ can be set, for instance, to $\epsilon = 0.01$. To solve this problem, we need a way to reckon the values of $\bar{k}(\kappa(k))$ from the relation $\kappa (k)$. To that end, it is useful to consider the probability for two nodes with hidden degrees $\kappa$ and $\kappa'$ to be connected in the ensemble of networks with global parameters $R = N/(2 \pi)$, $\mu$, $\beta$ and uniformly distributed angular coordinates. This probability is given by
\begin{align}\label{eq:conn_prob_ens}
  p(a_{\kappa\kappa'} = 1) & = \int \limits_{0}^{\pi} \frac{1}{\pi} \frac{1}{1 + \left( \frac{R \Delta \theta}{\mu \kappa \kappa'} \right)^{\beta}} \mathrm{d} \Delta \theta \nonumber \\
  & = {}_2 F_{1} \left( 1, \frac{1}{\beta}, 1 + \frac{1}{\beta}, -\left( \frac{R \pi}{\mu \kappa \kappa'}\right)^{\beta} \right)\ .
\end{align}
Starting from the initial guess $\kappa(k) = k$, we perform the following steps to refine the relation $\kappa(k)$:
\begin{itemize}
\item[1.] \textbf{Initialize expected degrees:} For every degree class $k$, set $\bar{k}(\kappa(k)) = 0$.
\item[2.] \textbf{Compute expected degrees:} For every pair of degree classes $(k,k')$, compute $P \equiv p(a_{\kappa(k) \kappa(k')} = 1)$ using Eq.~\eqref{eq:conn_prob_ens}. Set $\bar{k}(\kappa(k)) + N_{k'}P \rightarrow \bar{k}(\kappa(k))$ and $\bar{k}(\kappa(k')) + N_{k}P \rightarrow \bar{k}(\kappa(k'))$. By doing so, we add the expected number of connections of a node in degree class $k$ with nodes in degree class $k'$ and vice-versa. Notice that, when $k = k'$, we set $\bar{k}(\kappa(k))) + ( N_{k} - 1) P \rightarrow \bar{k}(\kappa(k))$ instead.
\item[3.] \textbf{Compute largest deviation:} Let $\epsilon_{\textrm{max}} = \max \lbrace | \bar{k}(\kappa(k)) - k | \rbrace_{k}$ be the maximal deviation between degrees and expected degrees. If $\epsilon_{\textrm{max}} > \epsilon$, the values of $\kappa(k)$ need to be corrected. Then, for every degree class $k$, set $| \kappa(k) + [k - \bar{k}(\kappa(k))]u | \rightarrow \kappa(k)$, where $u$ is a random variable drawn from $U(0,1)$. The rationale behind this transformation is that every degree-class hidden degree is corrected according to its expected-degree excess or deficiency; the random variable $u$ prevents the process from getting trapped in a local minimum. Next, go to step 1 to compute the expected degrees corresponding to the new set of $\kappa(k)$. Otherwise, if $\epsilon_{\textrm{max}} \leq \epsilon$, hidden degrees have been inferred for the current global parameters.
\end{itemize}

\subsection{Inferring parameter $\beta$}
To infer $\beta$, we need to compute the expected mean local clustering $ \bar{c}$ given the current values of the global parameters as well as of the hidden-degree distribution provided by $\kappa(k)$ and $N_{k}$ found using the algorithm from the last subsection. The method is based on the following idea. Suppose we want to estimate the expected clustering $\bar{c} (k)$ of some node with degree $k$. According to the definition of mean local clustering, this quantity is given by the probability for two randomly chosen neighbors of the node to be connected, which can be computed in two steps: first, we randomly choose two of its neighbors and draw their distances to the node from the distribution of distances between connected nodes in the model. Second, we compute the distance between the two neighbors and, with it, the probability for them to be connected. Two important points require some clarification:
\begin{itemize}
\item[a.] The model is uncorrelated at the hidden level. Therefore, in the calculation of the clustering, we draw the two neighbors from the uncorrelated distribution $P (k | k') = k P(k)/\langle k \rangle$.
\item[b.] The distribution of angular distance $\Delta \theta$ between two connected nodes with hidden degrees $\kappa$ and $\kappa'$, $\rho (\Delta \theta | a_{\kappa \kappa'} = 1)$, where $a_{\kappa \kappa'}$ stands for the corresponding adjacency-matrix element, is given by
\begin{equation}\label{eq:delt_theta_bayes}
\rho (\Delta \theta | a_{\kappa \kappa'} = 1) = \frac{p(a_{\kappa \kappa'} = 1|\Delta \theta) \rho(\Delta \theta)}{p(a_{\kappa \kappa'} = 1)}\ .
\end{equation}
In the above expression, $p(a_{\kappa \kappa'}=1|\Delta \theta) = 1/\left( 1 + \left( R \Delta \theta/(\mu \kappa \kappa') \right)^{\beta} \right)$ is the connection probability between the two nodes with hidden degrees $\kappa$ and $\kappa'$ separated by a distance $\Delta \theta$. The distribution of distances in the $\mathbb{S}^{1}$ model is simply $\rho(\Delta \theta) = 1/\pi$, since angular coordinates are uniformly distributed. Finally, $p(a_{\kappa \kappa'} = 1)$ is given in Eq.~\eqref{eq:conn_prob_ens}. Equation \eqref{eq:delt_theta_bayes} therefore reads
\begin{equation}\label{eq:delt_theta_expression}
\rho (\Delta \theta | a_{\kappa \kappa'} = 1) = \frac{\frac{1}{\pi} \frac{1}{1 + \left( \frac{R \Delta \theta}{\mu \kappa \kappa'} \right)^{\beta}}}{{}_2 F_{1} \left( 1, \frac{1}{\beta}, 1 + \frac{1}{\beta}, -\left( \frac{R \pi}{\mu \kappa \kappa'}\right)^{\beta} \right)}\ .
\end{equation}
\end{itemize}

The expected mean local clustering can now be found following three steps:
\begin{itemize}
\item[1.] \textbf{Initialize mean local clustering:} Let $ \bar{c} (k)$ represent the expected mean local clustering of degree class $k$. Set $\bar{c} (k) = 0$ for all $k$.
\item[2.] \textbf{Compute expected mean local clustering spectrum:} For every degree class $k$, do $m$ times:
\begin{itemize}
\item[i.] Draw two variables $k_{i}$ from $P(k_{i}|k), i=1,2$.
\item[ii.] Draw the corresponding random variables $\Delta \theta_{i}$ from the distributions $\rho (\Delta \theta_{i} | a_{\kappa(k) \kappa(k_{i})} = 1),i=1,2$ given in Eq.~\eqref{eq:delt_theta_expression}.
\item[iii.] Consider the two semicircles spanned by the diameter of the circle passing through the degree-$k$ node. It is equally likely for its two neighbours to lay in the same or in different semicircles. Hence, with probability $1/2$, set $\Delta \theta_{12} = \min ( | \Delta \theta_1 + \Delta \theta_2|, 2 \pi - | \Delta \theta_1 + \Delta \theta_2|)$ or $\Delta \theta_{12} = | \Delta \theta_1 - \Delta \theta_2|$.
\item[iv.] Set $\bar{c} (k) + p_{12}/m \rightarrow \bar{c}(k)$, where $p_{12} = \frac{1}{ 1 + \left( \frac{R \Delta \theta_{12}}{\mu \kappa(k_1) \kappa(k_2)} \right)^{\beta} }$ is the probability for nodes 1 and 2 to be connected.
\end{itemize}
\item[3.] \textbf{Compute expected mean local clustering:} The expected mean local clustering $\bar{c}$ can be readily computed as $\bar{c}  = \sum_{k} \bar{c} (k) N_{k}/N$.
\end{itemize}
If the error in the expected mean local clustering $| \bar{c} - \bar{c}^\mathrm{emp}| < \epsilon_{\bar{c}}$, where $\epsilon_{\bar{c}}$ is the desired precision and $\bar{c}^\mathrm{emp}$ is the observed mean local clustering coefficient, we can accept the current value of $\beta$ and proceed to the inference of the angular coordinates. Otherwise, $\beta$ needs to be corrected and the hidden degrees must be recalculated accordingly by repeating the process explained in the previous subsection. Notice that, since the expected mean local clustering coefficient is a monotonic function of $\beta$, the process can be iterated efficiently using the bisection method. In practice, however, we use a modified version of the bisection method in which the upper bound is let free until we reach a value of $\beta$ for which the expected clustering is higher than the observed one. More precisely, we start with a value for $\beta$ picked uniformly between 2 and 3. Then, while the expected clustering is lower than the observed one, we increase $\beta$ by multiplying it by 1.5. When we reach a value for which the observed clustering is surpassed, we start the regular bisection method. We also note that, for $\epsilon_{\bar{c}} = 0.01$, $m = 600$ is enough. Of course, if more precision is required, $m$ must be increased to guarantee that the fluctuations in the computed $ \bar{c}(k)$ are small enough.

\subsection{Angular coordinates}

Having inferred the values for the parameters $\beta$, $\mu$ and $\{\kappa_i\}$, we are in a position to infer the angular coordinates, $\{\theta_i\}$, of each node. This is performed by following two steps: a \textit{machine learning} step providing an initial ordering of the nodes as well as realistic positions, and a second step in which nodes are moved in order to maximize the likelihood that the $\mathbb{S}^1$ model generated the original edge list.

\subsection{Initial ordering and positions}

This step is a modified version of the Laplacian Eigenmaps (LE) algorithm introduced in Ref.~\cite{Belkin:2001} and used in Refs.~\cite{Alanis-Lobato:2016uq,muscoloni2017machine}. This machine learning algorithm was originally conceived for dimensionality reduction. The main idea is as follows. Given a set of points in $\mathbb{R}^{n}$, the algorithm first constructs a RGG by, for instance, connecting points located at a distance below some threshold in $n$-dimensional Euclidean space. Once this graph is known, the points are mapped to $\mathbb{R}^{m}$ with $m<n$ by diagonalizing the corresponding Laplacian and assigning to every point $i$ the coordinates $\mathbf{y}_{i} = (v_1^i, \cdots, v_m^i)$, where $v_j^i$ is the $i$-th component of the $j$-th Laplacian eigenvector with non-null eigenvalue (the eigenvectors are ordered according to their eigenvalues). It can be shown that these coordinates minimize the squared distances between connected pairs in the RGG,
\begin{equation}
\epsilon = \sum\limits_{i,j} \left| \mathbf{y}_i - \mathbf{y}_j \right|^2,
\end{equation}
while preventing all nodes from collapsing into a single point. Furthermore, the relevance of every connection $(i, j)$ in the RGG in the above expression can be modulated by assigning a weight $\omega_{ij}$ to it according to
\begin{equation}\label{eq:weights}
\omega_{ij} = e^{- \frac{\left| \mathbf{x}_i - \mathbf{x}_j \right|^2}{t}},
\end{equation}
where $\left| \mathbf{x}_i - \mathbf{x}_j \right|$ is the distance between the points in $\mathbb{R}^{n}$ and $t$ is a scaling factor fixed as the mean of the squares of all the distances~\cite{muscoloni2017machine}. The same procedure then leads to the minimization of
\begin{equation}
\epsilon = \sum\limits_{i,j} \left| \mathbf{y}_i - \mathbf{y}_j \right|^2 \omega_{ij},
\end{equation}

The approach taken by Refs.~\cite{Alanis-Lobato:2016uq,muscoloni2017machine} is to consider the network to be embedded as the RGG generated in a higher-dimensional space. Hence, by proceeding to the dimensionality reduction in $m$ (typically $m =2$) dimensions, we obtain an embedding in $R^{m}$, which can be radially normalized so that all points lay in $\mathbb{S}^{m-1}$. The improvement in Ref.~\cite{muscoloni2017machine} is to assign weights according to some heuristic and, once the coordinates on the plane are known, these are used to infer the ordering of nodes only. The final coordinates are computed by distributing all nodes on the circle with $\theta_{i+1} - \theta_{i} = 2 \pi / N, \, \forall i$. We now propose an improvement based on assigning the weights in the Laplacian matrix as well as the gaps $\theta_{i+1} - \theta_{i}$ according to the $\mathbb{S}^1$ model.

\begin{itemize}
\item[1.] \textbf{Laplacian Eigenmaps for node ordering:} Since degree-one nodes do not add geometric information, we remove them at this step and work with the subgraph of nodes with $k > 1$. We then apply the Laplacian Eigenmaps method to such graph after weighting every links according to Eq.~\eqref{eq:weights}, where we use
\begin{equation}
\left| \mathbf{x}_i - \mathbf{x}_j \right| = 2 \sin \frac{ \langle \Delta \theta_{ij} \rangle}{2}
\end{equation}
as a proxy for the distance $\left| \mathbf{x}_i - \mathbf{x}_j \right|$ and $\langle \Delta \theta_{ij} \rangle$ is the expected angular distance between nodes $i$ and $j$ in the $\mathbb{S}^{1}$ model conditioned to the fact that they are connected. The above expression simply maps such expected angular distance onto the corresponding cord length, since Laplacian Eigenmaps is designed to work on Euclidean space and, therefore, it seems natural to consider the $\mathbb{S}^{1}$ model as embedded in $\mathbb{R}^{2}$ for the algorithm. The expected distance between the nodes can be readily computed from Eq.~\eqref{eq:delt_theta_expression} as
\begin{align}
  \langle \Delta \theta_{ij} \rangle & = \int \limits_{0}^{\pi} \Delta \theta_{ij} \rho (\Delta \theta_{ij} | a_{ij} = 1) \mathrm{d} \Delta \theta_{ij} \nonumber \\
  &= \frac{\pi\ {}_2 F_{1} \left( 1, \frac{2}{\beta}, 1 + \frac{2}{\beta}, -\left( \frac{R \pi}{\mu \kappa(k_i) \kappa(k_j)}\right)^{\beta} \right)}{2\ {}_2 F_{1} \left( 1, \frac{1}{\beta}, 1 + \frac{1}{\beta}, -\left( \frac{R \pi}{\mu \kappa(k_i) \kappa(k_j)}\right)^{\beta} \right)} \ .
\end{align}
Since a similar approach developed in Ref.~\cite{muscoloni2017machine} has been shown to yield very good results in terms of the angular ordering of the nodes, 
we use this machine learning step to define a sequence of angular coordinates
\begin{equation}
S' = ( \theta_1, \ldots, \theta_{N'} )
\end{equation}
for the nodes in the subgraph, where the angles in $S'$ are ordered in increasing order. Each $\theta_{i}$ is computed as
\begin{equation}
\theta_i = \mathrm{atan2} \left( v_2^i, v_1^i \right),
\end{equation}
where $v_1^i$ and $v_2^i$ are the $x$ and $y$ coordinates of node $i$ found by LE. Once we have the ordering of nodes with $k>1$, we reincorporate the degree-one nodes. This can be easily done by replacing every node $i$ with $t$ degree-one neighbors by the sequence $(n_{1}^i, \ldots, n_{\left\lfloor t/2 \right\rfloor}^{i}, i, n_{\left\lfloor t/2 \right\rfloor+1}^{i}, \ldots, n_{t}^{i})$ in $S'$, where $n_{j}^{i}$ is the $j$-th degree-one neighbor of node $i$ (in any arbitrary order) and $\lfloor \cdot \rfloor$ is the floor function. Such operation yields a new sequence of nodes $S$ including all the nodes in the original graph.

\item[2.] \textbf{Order-preserving adjustment:} This last step of the approximate embedding locates the nodes on the circle preserving the ordering of the nodes in $S$. To that end, we set every node's coordinate such that the gap between two consecutive nodes in $S$ is proportional to the expected gap between two consecutive nodes with the same hidden variables and adjacency-matrix element in the $\mathbb{S}^{1}$ model. We do this in two steps:

\begin{itemize}
\item[i.] \textit{Computing the expected gaps:} Let nodes $i$ and $i+1$ be consecutive in $S$. The distribution for the length of the gap $g_i$ between them can be obtained from Bayes' rule, as in Eq.~\eqref{eq:delt_theta_bayes},
\begin{equation}
\rho (g_i | a_{i+1,i}) = \frac{p(a_{i+1,i}|g_i) \rho(g_i)}{\int \limits_{0}^{\pi} p(a_{i+1,i}|g_i) \rho(g_i) \mathrm{d}g_i},
\end{equation}
where now $\rho(g_i)$ is an exponential distribution with mean $2 \pi / N$,
\begin{equation}
\rho(g_i) = \frac{N}{2 \pi} e^{-\frac{N}{2 \pi}g_i},
\end{equation}
and
\begin{align}
  p(a_{i+1,i}|&g_i) = \left( \frac{1}{1 + \left( \frac{R g_i}{\mu \kappa(k_{i+1}) \kappa(k_i)} \right)^{\beta}} \right)^{a_{i+1,i}} \nonumber \\
  & \times \left( 1 - \frac{1}{1 + \left( \frac{R g_i}{\mu \kappa(k_{i+1}) \kappa(k_i)} \right)^{\beta}} \right)^{1- a_{i+1,i}} \ .
\end{align}
The expected gap $\langle g_i \rangle$ can thus be computed as
\begin{equation}
\langle g_i \rangle = \frac{\int \limits_{0}^{\pi} g_i p(a_{i+1,i}|g_i) \rho(g_i) \mathrm{d}g_i}{\int \limits_{0}^{\pi} p(a_{i+1,i}|g_i) \rho(g_i) \mathrm{d}g_i}\ .
\end{equation}
Both integrals can be carried out numerically.

\item[ii.] \textit{Normalizing the gaps:} By applying the last step to every pair of consecutive nodes (including the pair $(N,1)$), we obtain a sequence of $N$ expected gaps which, however, needs not sum up to $2 \pi$, so we normalize each $\langle g_i \rangle$ as
\begin{equation}
\tilde{g}_i = 2\pi \frac{\langle g_i \rangle }{\sum\limits_{i=1}^{N} \langle g_i \rangle }\ .
\end{equation}
Finally, we can assign every node's coordinate sequentially, starting with $\theta_1 = 0$, as $\theta_i = \theta_{i-1} + \tilde{g}_{i-1}, \, i = 2, \ldots, N$.
\end{itemize}

\end{itemize}

\subsection{Likelihood maximization}
\label{maxlike}
This stage of the embedding algorithm adjusts the angular coordinates that maximize the likelihood for the observed network to be generated by the model. As opposed to previously proposed likelihood-maximization schemes, we do not need to explore a vast region of configuration space, since the machine learning stage provides a set of coordinates located near an optimal configuration. Hence, we visit every node once and propose several new angular coordinates for it, keeping the one with higher log-likelihood. The steps we follow are:

\begin{itemize}
\item[1.] \textbf{Define a new ordering of nodes:} We visit the nodes in the order defined by the network's onion decomposition. In the sequence, the ordering among nodes belonging to the same layer in the decomposition is random.

\item[2.] \textbf{Find new optimal coordinates:} For every node $i$, we select the optimal coordinate among candidate positions generated in the vicinity of the mean angular coordinate of its neighbors. This is achieved in three steps:

\begin{itemize}
\item[i.] \textit{Compute mean coordinate of node $i$'s neighbors:} Let node $i$ have $k_i$ neighbors, which we now label with index $j = 1, \ldots, k_i$. Since nodes lay on a circle, we must compute their mean angular coordinate $\bar{\theta}_i$ using the vector sum of their positioning vectors in $\mathbb{R}^2$. The polar angle of the resulting vector sum is given by
\begin{equation}
\bar{\theta}_i = \mathrm{atan2} \left( \sum_{j} \frac{1}{\kappa_j^2}\sin \theta_j, \sum_{j} \frac{1}{\kappa_j^2}\cos \theta_j  \right)\ ,
\end{equation}
where the hidden degrees in the above expression weight the contribution of every neighbor's positioning vector, as proposed in Ref.~\cite{blasius2016efficient}.

\item[2.] \textit{Propose new positions around $\bar{\theta}_i$:} We generate $100 \max ( \ln N, 1 )$ candidate angular coordinates from a normal distribution with mean $\bar{\theta}_i$ and standard deviation $\sigma$ given by
\begin{equation}
\sigma = \max \left( \frac{\pi}{12}, \frac{\Delta \theta_{\mathrm{max}}}{2} \right)\ ,
\end{equation}
where $\Delta \theta_{\mathrm{max}}$ is the angular distance between $\bar{\theta}_i$ and the most distant neighbor of node $i$, i.e.,
\begin{equation}
 \Delta \theta_{\mathrm{max}} = \max \lbrace \min \left( | \theta_j - \bar{\theta}_i |, 2 \pi - | \theta_j - \bar{\theta}_i |  \right) \rbrace_{j}\ .
\end{equation}

\item[3.] \textit{Select the most likely candidate position:} Compute the local log-likelihood of every candidate position as well as of node $i$'s current angular coordinate according to
\begin{equation}
\ln \mathcal{L}_i = \sum \limits_{j \neq i} a_{ij} \ln p_{ij} + (1 - a_{ij} ) \ln (1 - p_{ij})\ .
\end{equation}
Locate node $i$ at the angular position maximizing the local log-likelihood.

\end{itemize}
\end{itemize}

\subsection{Adjusting hidden degrees}\label{sec:adjustment_hidden_degrees}

The final process adjusts hidden degrees according to the hidden coordinates found so that $\bar{k}(\kappa_i) = k_i$. The algorithm is similar to the initial inference of hidden degrees:

\begin{itemize}
\item[1.] \textbf{Compute expected degrees:} For every node $i$, set
\begin{equation}
\bar{k}(\kappa_i) = \sum \limits_{j \neq i} \frac{1}{1 + \left( \frac{R \Delta \theta_{ij}}{\mu \kappa_i \kappa_j} \right)^{\beta}}.
\end{equation}

\item[2.] \textbf{Correct hidden degrees:} Let $\epsilon_{\textrm{max}} = \max \lbrace | \bar{k}(\kappa_i) - k_i | \rbrace_{i}$ be the maximal deviation between degrees and expected degrees. If $\epsilon_{\textrm{max}} > \epsilon$, the set of hidden degrees needs to be corrected. Then, for every node $i$, set $| \kappa_i + [k_i - \bar{k}(\kappa_i)]u | \rightarrow \kappa_i$, where $u$ is a random variable drawn from $U(0,1)$. As in Sec.\ref{sec:inf_hid_degs}, the random variable $u$ prevents the process from getting trapped in a local minimum. Next, go to step 1 to compute the expected degrees corresponding to the new set of hidden degrees. Otherwise, if $\epsilon_{\textrm{max}} \leq \epsilon$, hidden degrees have been inferred for the current global parameters and angular coordinates.
\end{itemize}


%

\end{document}